\documentstyle[twoside,fleqn,espcrc2,epsfig]{article}

\def\Journal#1#2#3#4{{#1} {\bf #2}, #3 (#4)}

\def\NPB{{\em Nucl. Phys.} B}
\def\NPPS{\em Nucl. Phys. (Proc. Suppl.)} 
\def\PLB{{\em Phys. Lett.}  B}
\def\PRL{\em Phys. Rev. Lett.}
\def\PRD{{\em Phys. Rev.} D}
\def\ZPC{{\em Z. Phys.} C}

\def\be{\begin{equation}}
\def\ee{\end{equation}}
\def\bea{\begin{eqnarray}}
\def\eea{\end{eqnarray}}
\newcommand{\lwig}{\mbox{\,\raisebox{.3ex}
    {$<$}$\!\!\!\!\!$\raisebox{-.9ex}{$\sim$}\,}}
\newcommand{\gwig}{\mbox{\,\raisebox{.3ex}
    {$>$}$\!\!\!\!\!$\raisebox{-.9ex}{$\sim$}}\,}

\newcommand{\ai}{{\overline{I}}}
\newcommand{\iai}{I\overline{I}} 
\newcommand{\xpr}{{x^\prime}}

\begin{document}

%\pagestyle{empty}

% declarations for front matter

\title{\vspace{-3.5cm}
\begin{flushleft}
{\normalsize DESY 98--114}\\
{\normalsize hep-ph/9808422}
\end{flushleft}
       \vspace{1cm}
Searching for QCD--Instantons at 
      HERA%
\thanks{Talk presented at the QCD Euroconference 98, 
        Montpellier, July~2--8, 1998; to be published in the
        Proceedings.}
      }

\author{A. Ringwald\address{DESY, Notkestr. 85, D-22603 Hamburg, Germany}
        \addtocounter{address}{-1} 
        and F. Schrempp\addressmark}

\begin{abstract}
We review the present status of our ongoing systematic study
of the discovery potential of QCD-instanton induced events in 
deep-inelastic scattering at HERA.
\end{abstract}

\maketitle

\section{INTRODUCTION}\label{s0}

Instantons~\cite{bpst}, %non-perturbative 
 fluctuations of non-abelian gau\-ge 
fields representing topology changing tunnelling transitions in Yang-Mills 
gauge 
theories, induce hard processes which are absent in conventional perturbation 
theory~\cite{th}: In accord with the ABJ-anomaly, they violate certain 
fermionic quantum numbers, notably, chirality ($Q_{5}$) in (massless) QCD 
and baryon plus lepton number ($B+L$) in electro-weak interactions. 

While implications of QCD-instantons, notably for long-distance phenomena, 
have been intensively studied
for a long time, mainly in the context of the phenomenological instanton 
liquid model~\cite{ss} and of lattice simulations~\cite{lattice}, the 
direct experimental verification of their existence is
lacking up to now. Clearly, an experimental discovery of such a novel, 
non-perturbative manifestation of non-abelian gauge theories would 
be of basic significance. 

The deep-inelastic regime is distinguished by the fact that here
hard QCD-in\-stan\-ton induced processes may both be {\it
calculated}~\cite{bb,mrs1,rs-pl} within in\-stan\-ton-per\-tur\-ba\-tion
theory  and possibly 
{\it detected experimentally}~\cite{rs,grs,dis97-phen,crs}.

In this paper, we review the present status of our ongoing systematic
study~\cite{mrs1,rs-pl,rs,grs,dis97-phen,crs} of the discovery potential
of QCD-instanton induced events in deep-inelastic scattering (DIS) at HERA.  

\section{CROSS-SECTION ESTIMATES}\label{s1}

\begin{figure}
\vspace{-0.25cm}
\begin{center}
\epsfig{file=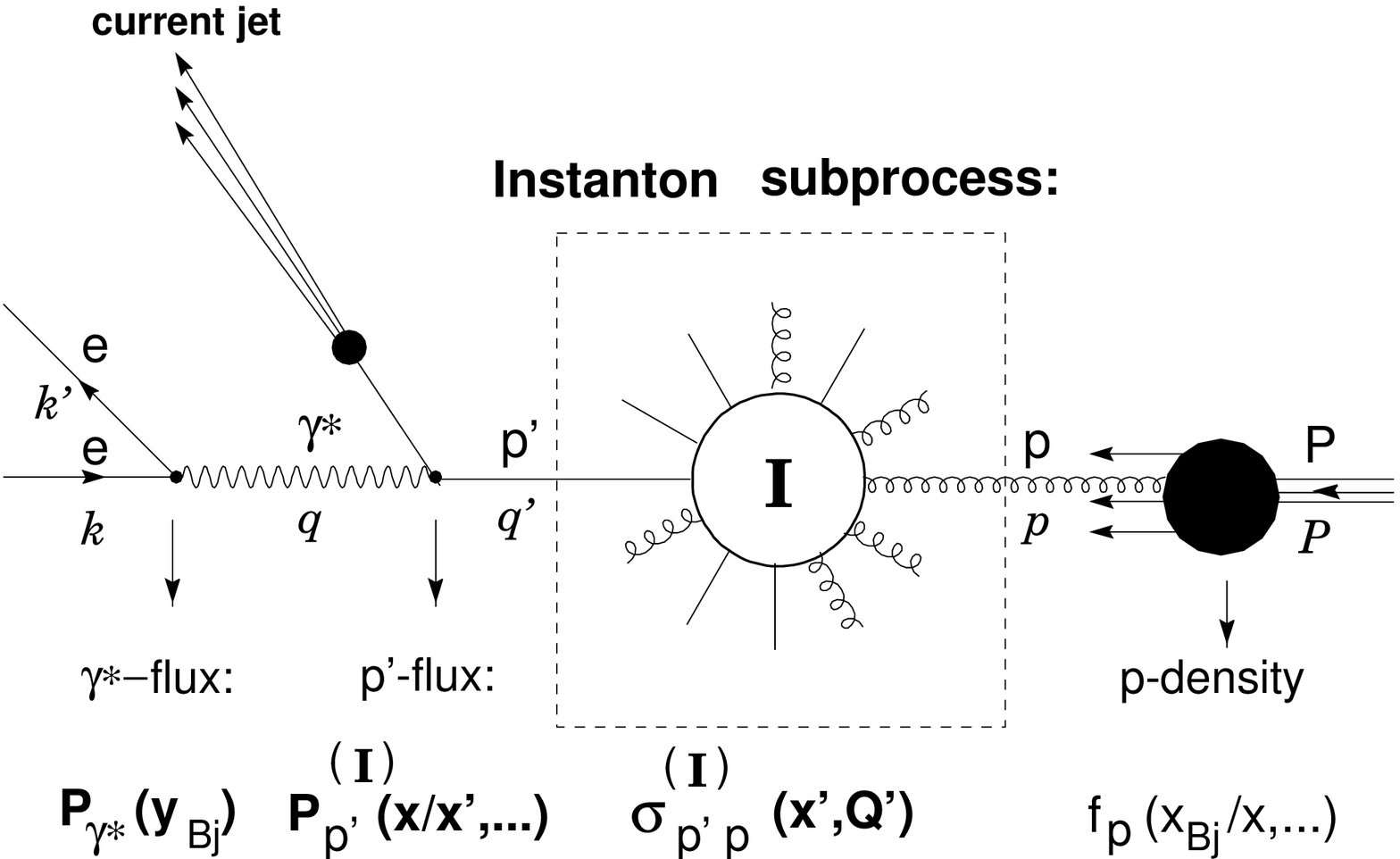,width=7cm}
\vspace{-0.5cm}
\caption[dum]{\label{ev-displ}The leading instanton-induced process 
in the DIS regime of $e^\pm P$ scattering, violating chirality by
$\triangle Q_5=2n_f$.}
\vspace{-0.8cm}
\end{center}
\end{figure}

The leading instanton ($I$)-induced process in the DIS regime of $e^\pm
P$ scattering is displayed in Fig.~\ref{ev-displ}. 
The dashed box emphasizes the  so-called  instanton-{\it subprocess}
with its own Bjorken variables,
        \begin{equation}
        Q^{\prime\,2}=-q^{\prime\,2}\geq 0;\hspace{0.5cm}
        \xpr=\frac{Q^{\prime\,2}}{2 p\cdot q^\prime}\le 1.
        \end{equation}         
%%%%%%%%%%%%%%%%%%%%%%%%%%%%%FIGURE  %%%%%%%%%%%%%%%%%%%%%%%%%%%%%%%%%
%%%%%%%%%%%%%%%%%%%%%%%%%%%%%%%%%%%%%%%%%%%%%%%%%%%%%%%%%%%%%%%%%%%%%%
The inclusive $I$-induced cross-section in unpolarized deep-inelastic 
$e^\pm P$ scattering can be expressed (in the Bjorken limit) as~\cite{rs-pl}
\begin{equation}
\label{ePcross}
\frac{d\sigma_{eP}^{(I)}}{d\xpr\,dQ^{\prime 2}} \simeq
\sum_{p^\prime,p}
\frac{d{\mathcal L}^{(I)}_{p^\prime p}}{d\xpr\,dQ^{\prime 2}}
\,\sigma_{p^\prime p}^{(I)}(\xpr,Q^{\prime 2}) ,
\end{equation} 
where $p^\prime =q^\prime,\overline{q}^\prime$ denotes the virtual 
quarks entering the $I$-subprocess, with corresponding total cross-section
$\sigma_{p^\prime p}^{(I)}$, from the photon side and
$p=q,\overline{q},g$ denotes the target partons. 
The differential luminosity $d{\mathcal L}^{(I)}_{p^\prime p}$, accounting
for the number of $p^\prime p$ collisions per $eP$ collision, has a 
convolution-like structure~\cite{rs}, involving integrations over the 
target-parton density, $f_p$, the $\gamma^\ast$-flux,
$P_{\gamma^\ast}$, and the {\it known}~\cite{dis97-phen} 
flux $P^{(I)}_{p^\prime}$ of the parton $p^\prime$ in the 
$I$-background. 

In Eq.~(\ref{ePcross}), the $I$-subprocess
total cross-section $\sigma_{p^\prime p}^{(I)}$ 
contains the essential instanton dynamics. We have 
evaluated the latter~\cite{rs-pl} 
by means of the optical theorem and the so-called  
$\iai$-valley approximation~\cite{valley} for the relevant 
$q^\prime g\Rightarrow q^\prime g$ forward elastic scattering
amplitude in the $\iai$ background. This method 
resums the exponentiating final state gluons in form of the known
valley action $S^{(\iai)}$ and reproduces standard $I$-perturbation theory at
larger  $\iai$ separation $\sqrt{R^2}$. 

Corresponding to the symmetries of the theory, the instanton calculus
introduces at the classical level certain (undetermined) 
``collective coordinates'' like the $I\,(\overline{I})$-size parameters 
$\rho\,(\overline{\rho})$ and the $\iai$ distance 
$\sqrt{R^2/\rho\overline{\rho}}$ (in units of the size).  
Observables like $\sigma_{p^\prime p}^{(I)}$
must be independent thereof and thus involve integrations over all collective
coordinates. Hence, we have generically,
\begin{eqnarray}
\label{sigma}
        \sigma_{p^\prime p}^{(I)}
        &=&\int\limits_0^\infty {\rm d}\rho\,D(\rho)
        \int\limits_0^\infty {\rm d}\overline{\rho}\, 
        D(\overline{\rho})
        \int {\rm d}^4 R\ \ldots        
        \\
\nonumber
        &&\times\,
        {\rm e}^{-({\rho + \overline{\rho}})Q^\prime}
        {\rm e}^{i(p+q^\prime )\cdot R}
        {\rm e}^{-\frac{4\pi}{\alpha_s}(S^{(\iai)}(\xi)-1)}. 
\end{eqnarray}   

The first important quantity of interest, entering Eq.\,(\ref{sigma}), is 
the $I$-density, $D(\rho)$ (tunnelling amplitude).
It has been worked out a long time ago~\cite{th,morretal} in the framework of 
$I$-perturbation theory: (renormalization scale $\mu_r$)
\begin{eqnarray}
   D(\rho)
   &=&
   d \left(\frac{2\pi}{\alpha_s(\mu_r)}\right)^6
   \exp{(-\frac{2\pi}{\alpha_s(\mu_r)})}\frac{(\rho\, \mu_r)^b}{\rho^{\,5}},
   \label{density}
\\ 
\label{b-morretal}
b&=&\beta_0+\frac{\alpha_s(\mu_r)}{4\pi}
      (\beta_1-12\beta_0) ,
\end{eqnarray}
in terms of the QCD 
$\beta$-function coefficients, $\beta_0=11-\frac{2}{3}{n_f},\ 
\beta_1=102-\frac{38}{3} {n_f}$. In this form it satisfies 
renormalization-group invariance at the two-loop level~\cite{morretal}.
Note that the large, positive power $b$ of $\rho$ in the 
$I$-density~(\ref{density}) would make the integrations over the 
$I(\ai)$-sizes in Eq.~(\ref{sigma}) infrared divergent without the  
crucial exponential cut-off~\cite{mrs1} 
$e^{-(\rho+\overline{\rho})\,Q^\prime}$ arising from the virtual quark
entering the $I$-subprocess from the photon side. 

The second important quantity of interest, entering Eq.\,(\ref{sigma}), is 
the $\iai$-interaction, $S^{(\iai)}-1$. In the valley approximation, the 
$\iai$-valley action, 
$S^{(\iai)}\equiv \frac{\alpha_s}{4\pi}\,S[A_\mu^{(\iai)}]$, 
is restricted by conformal
invariance to depend only on the ``conformal separation'', 
$\xi=R^2/\rho\overline{\rho}+\rho/\overline{\rho}+\overline{\rho}/\rho$,
and its functional form is explicitly known~\cite{valley}. It is important
to note that, for all separations $\xi$, the interaction between $I$
and $\ai$ is {\it attractive}; in particular, the $\iai$-valley action 
monotonically decreases from 1 at infinite conformal separation to 
0 at $\xi =2$, corresponding to $R^2=0,\rho =\overline{\rho}$. 

The collective coordinate integration in the cross-section~(\ref{sigma})
can be performed via {\it saddle-point techniques}. One finds 
$R_\mu^\ast = (\rho^\ast \sqrt{\xi^\ast -2},\vec{0})$ and 
$\rho^\ast=\overline{\rho}^\ast$, where the saddle-point solutions $\rho^\ast$
and $\xi^\ast$ behave qualitatively as
\begin{equation}
\label{qual-exp}
\rho^\ast \sim \frac{4\pi}{\alpha_s Q^\prime};\hspace{4ex}
\sqrt{\xi^\ast -2} = \frac{R^\ast}{\rho^\ast} 
\sim 2\, \sqrt{\frac{\xpr}{1 -\xpr}} .
\end{equation}
Thus, the virtuality $Q^\prime$ controls the effective $I(\ai)$-size:  
as one might have expected intuitively, highly virtual quarks probe only
small instantons. 
The Bjorken-variable $\xpr$, on
the other hand, controls the conformal separation between $I$ and $\ai$:
for decreasing $\xpr$, the conformal separation decreases.

%%%%%%%%%%%%%%%%%%%%%%%%%%%%%FIGURE  %%%%%%%%%%%%%%%%%%%%%%%%%%%%%%%%%%%5
\begin{figure}[t]
\vspace{-0.65cm}
\begin{center}
\epsfig{file=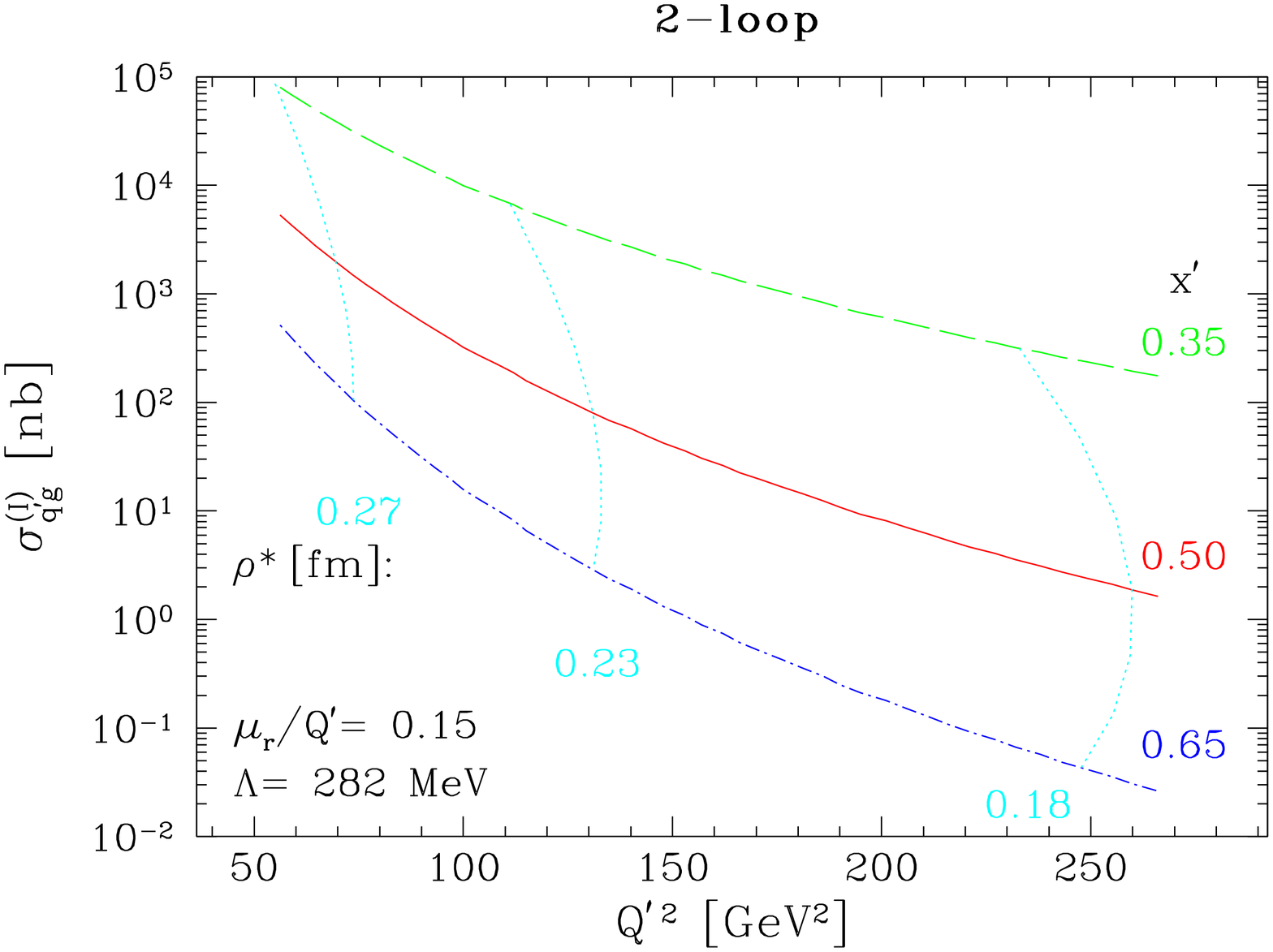,angle=-90,width=7cm}
\epsfig{file=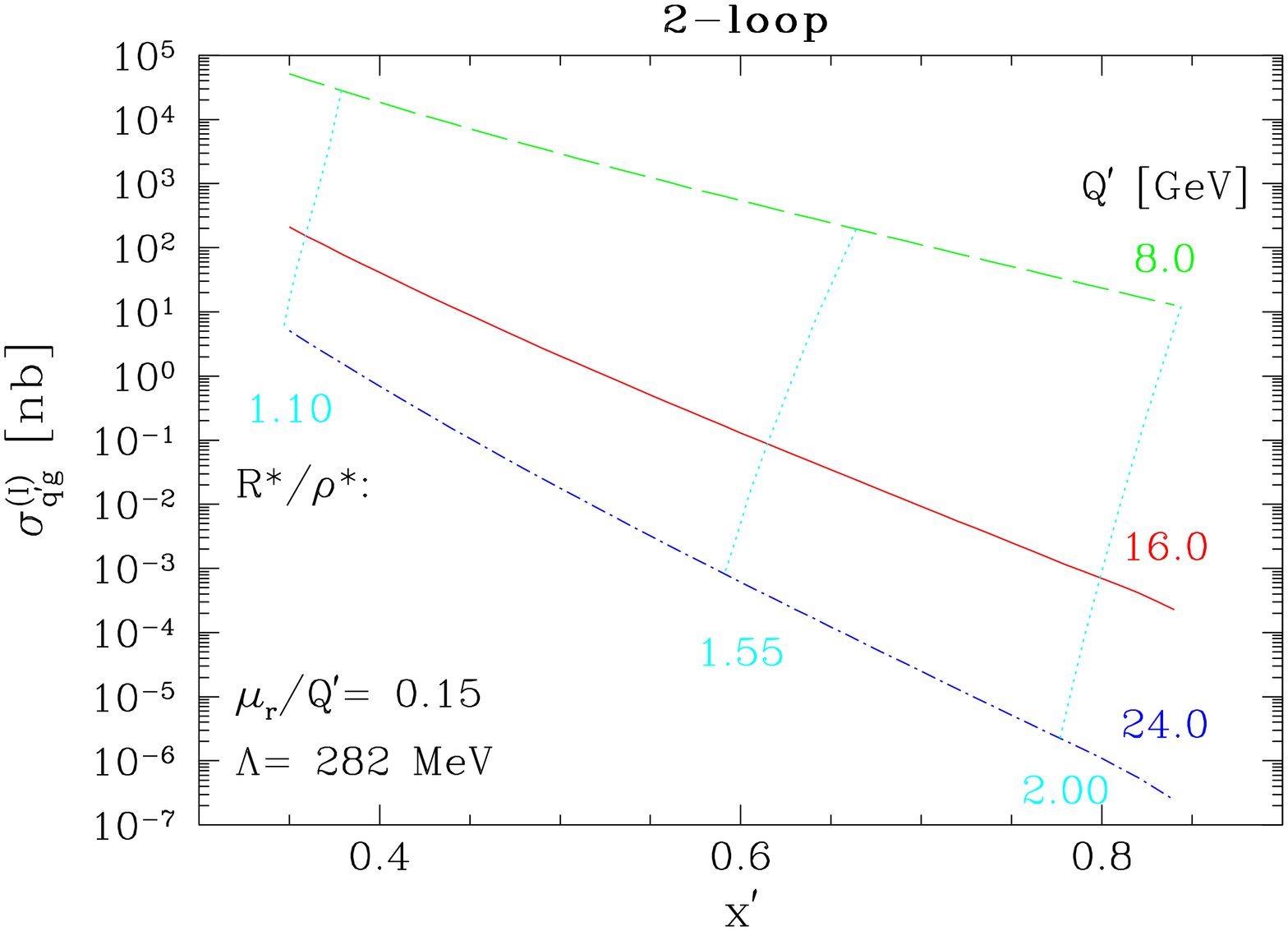,angle=-90,width=7cm}
\vspace{-1.0cm}
\caption[dum]{\label{isorho}$I$-subprocess cross-section~\cite{rs-pl}.}
\vspace{-1cm}
\end{center}
\end{figure}
%%%%%%%%%%%%%%%%%%%%%%%%%%%%%%%%%%%%%%%%%%%%%%%%%%%%%%%%%%%%%%%%%%%%%%
Our quantitative results~\cite{rs-pl} on the {\it dominating} cross-section for
a target gluon, $\sigma^{(I)}_{q^\prime g}$, are shown in detail 
in Fig.~\ref{isorho}, both as functions of $Q^{\prime 2}$ (top) and of $\xpr$
(bottom). The residual dependence on the renormalization scale turns 
out~\cite{rs-pl} to be strongly reduced by using the two-loop 
renormalization-group invariant form of the $I$-density $D(\rho)$ from 
Eqs.~(\ref{density}) and (\ref{b-morretal}).   
Intuitivelely one may expect~\cite{mrs1,bb} $\mu_r \sim 1/\langle \rho
\rangle \sim Q^\prime/\beta_0 ={\cal O}(0.1)\, Q^\prime$. Indeed, 
this guess turns out to match quite well
our actual choice of the  
 ``best'' scale, $\mu_r = 0.15\ Q^\prime$, determined by  
$\partial \sigma^{(I)}_{q^\prime g}/\partial \mu_r \simeq 0$. 
The dotted curves in Fig.~\ref{isorho}, indicating lines of constant 
$\rho^\ast$ (top) and of constant $R^\ast/\rho^\ast$ (bottom),    
nicely illustrate the qualitative relations~(\ref{qual-exp}) and their
consequences: the $Q^\prime$ dependence essentially maps the $I$-density, 
whereas the $\xpr$ dependence mainly maps the $\iai$-interaction.    

Fortunately, important information about the range of validity of 
$I$-perturbation for the $I$-density
and the $\iai$-interaction, in terms of the instanton collective coordinates 
($\rho\leq \rho_{\rm max}, R/\rho\geq (R/\rho)_{\rm min}$), can be obtained 
from recent (non-perturbative) lattice simulations of QCD and translated 
via the saddle-point relations~(\ref{qual-exp}) into a ``fiducial'' 
kinematical region 
($Q^\prime\geq Q^\prime_{\rm min},\xpr\geq x^\prime_{\rm min}$). 
In fact, from a comparison of the perturbative expression of the 
$I$-density~(\ref{density}) with recent lattice ``data''~\cite{ukqcd} 
one infers~\cite{rs-pl} semi-classical $I$-perturbation theory to be valid 
for $\rho \lwig \rho_{\rm max}\simeq 0.3$ fm.     
Similarly, it is found~\cite{rs-pl} that the attractive, semi-classical valley
result for the $\iai$-interaction applies down to a minimum conformal 
separation $\xi_{\rm min}\simeq 3$, corresponding to 
$(R^\ast/\rho^\ast)_{\rm min}\simeq 1$. The corresponding 
``fiducial'' kinematical region for our 
cross-section predictions in DIS is then obtained as
\begin{equation}
 \left.\begin{array}{lcl}\rho^\ast&\lwig&  
         0.3 {\rm\ fm};\\[1ex]
 \frac{R^\ast}{\rho^\ast}&\gwig&
  1\\
 \end{array}\right\}\Rightarrow
 \left\{\begin{array}{lclcl}Q^\prime&\geq &Q^\prime_{\rm min}&\simeq&
 8 {\rm\ GeV};\\[1ex]
 x^\prime&\geq &x^\prime_{\rm min}&\simeq &0.35.\\
 \end{array} \right .
\label{fiducial}
\end{equation}

%%%%%%%%%%%%%%%%%%%%%%%%%%%%%FIGURE  %%%%%%%%%%%%%%%%%%%%%%%%%%%%%%%%%%%5
\begin{figure}[t]
\vspace{-0.95cm}
\begin{center}
\epsfig{file=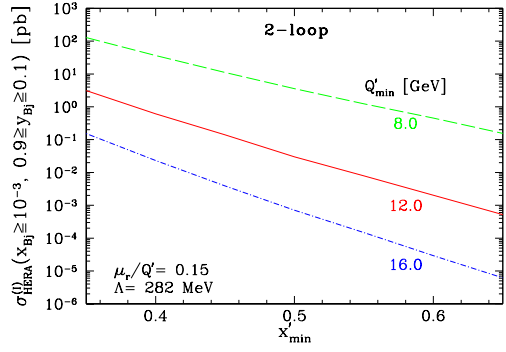,angle=-90,width=7.5cm}
\vspace{-1.0cm}
\caption[dum]{\label{sigHERA}$I$-induced cross-section at HERA~\cite{rs-pl}.}
\vspace{-1cm}
\end{center}
\end{figure}
%%%%%%%%%%%%%%%%%%%%%%%%%%%%%%%%%%%%%%%%%%%%%%%%%%%%%%%%%%%%%%%%%%%%%%
Fig.~\ref{sigHERA} displays the finalized $I$-induced cross-section at HERA,
as function of the cuts $x^\prime_{\rm min}$ and
$Q^\prime_{\rm min}$, as
obtained with the new release ``QCDINS 1.6.0''~\cite{crs} of our $I$-event
generator. For the following ``standard cuts'', 
\begin{eqnarray}
\label{standard-cuts} 
{\mathcal C}_{\rm std}&=&
\xpr\ge0.35,Q^\prime\ge 8\, {\rm GeV}, x_{\rm Bj}\geq 10^{-3},
\\ \nonumber
&& 0.1\leq y_{\rm Bj}\leq 0.9 ,
\end{eqnarray}
involving the minimal cuts~(\ref{fiducial}) extracted from lattice 
simulations, we specifically obtain 
\begin{eqnarray}
\label{minimal-cuts}
\sigma^{(I)}_{\rm HERA}({\mathcal C}_{\rm std}) 
&=& {\mathcal O}(100)\, {\rm pb}.
\end{eqnarray}
The main inherent uncertainties are discussed in Ref.~\cite{rs-pl}.
With the total luminosity accumulated by experiments at HERA, 
${\mathcal L}={\mathcal O}(80)$ pb$^{-1}$, there should be already 
${\mathcal O}(10^4)$ $I$-induced events from the kinematical 
region~(\ref{standard-cuts})
on tape. Note also that the cross-section quoted in Eq.~(\ref{minimal-cuts})
corresponds to a fraction of $I$-induced to normal DIS events of
$f^{(I)}({\mathcal C}_{\rm std}) = {\mathcal O}(1)\, \%$. 

\section{SEARCHES AT HERA}\label{s2}

Thus, it seems to be a question of {\it signature} rather than a 
question of {\it rate} to discover $I$-induced scattering processes at HERA. 
Hence, we turn now to the {\it final states} of $I$-induced events in DIS.

%%%%%%%%%%%%%%%%%%%%%%%%%%%%%FIGURE  %%%%%%%%%%%%%%%%%%%%%%%%%%%%%%%%%%%
\begin{figure}[t]
\begin{center}
\epsfig{file=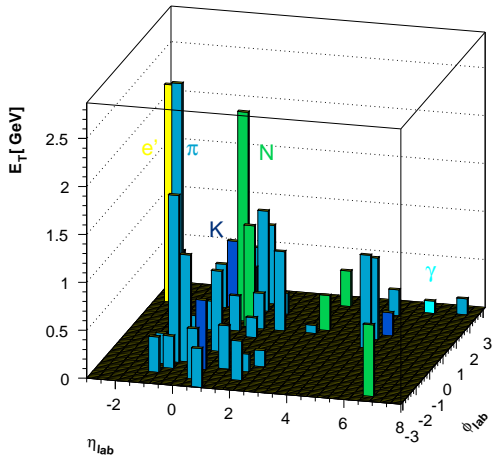,width=6.5cm}
\vspace{-0.5cm}
\caption[dum]{\label{typical}Lego plot of a typical $I$-induced event
in the HERA lab system.}
\vspace{-1cm}
\end{center}
\end{figure}
%%%%%%%%%%%%%%%%%%%%%%%%%%%%%%%%%%%%%%%%%%%%%%%%%%%%%%%%%%%%%%%%%%%%%%
In Fig.~\ref{typical} we display the lego plot of a typical $I$-induced event 
at HERA, as generated by our Monte-Carlo generator 
QCDINS~\cite{grs,dis97-phen,crs}. Its characteristics can be easily 
understood on the basis of the underlying $I$-subprocess: 

The current quark in Fig.~\ref{ev-displ} gives rise, after 
hadronization, to a current-quark jet. The partons from the $I$-subprocess, 
on the other hand, are emitted spherically symmetric in the   
$p^{\prime}p$ c.m. system. The gluon multiplicities are 
generated according to a Poisson distribution with mean multiplicity
$\langle n_g \rangle^{{(I)}}\sim 1/\alpha_s\sim 3$. 
The total mean parton multiplicity is large, of the order of ten. 
After hadronization we therefore expect from the $I$-subprocess a
final state structure reminiscent of a decaying fireball: Production
of the order of 20 hadrons, always containing strange mesons, concentrated in 
a ``band'' at fixed pseudorapidity $\eta$ in
the ($\eta$, azimuth angle $\phi$)-plane. Due to the boost 
from the $p^{\prime}p$ c.m. system to the HERA-lab system, the center of the
band is shifted in $\eta$ away from zero, and its width is 
of order $\Delta\eta\simeq 1.8$, as typical for a spherically symmetric event. 
The total invariant mass of the 
$I$-system, $\sqrt{s^{\prime}}=Q^{\prime}\sqrt{1/\xpr -1}$, 
is expected to be in the 10 GeV range, for $\xpr\simeq 0.35$, 
$Q^{\prime}\simeq 8$ GeV. All these expectations are clearly
reproduced by our Monte-Carlo simulation.

These features have been exploited by experimentalists at HERA to place 
first upper limits on the fraction of $I$-induced events to normal DIS 
(nDIS) events, in a similar kinematical region as our standard 
cuts~(\ref{standard-cuts}): From the search of a $K^0$ excess in the ``band'' 
region the H1 Collaboration could establish a limit of 
$f_{\rm lim}^{(I)}=6$ \%, while the search of an excess in charged 
multiplicity yields $f_{\rm lim}^{(I)}=2.7$ \%~\cite{h1-limits}.
The limit from the charged multiplicity distribution has been further 
improved in Ref.~\cite{ck-limits} to about 1 \%. 

Thus, despite of the high rate of $I$-induced events at HERA, no 
{\it single} observable is known (yet) with sufficient nDIS rejection.
A dedicated {\it multi}-observable analysis is required.
However, it seems that a {\it decisive} search for $I$-induced 
events at HERA is feasible.

\end{document}